\documentstyle[12pt,epsf,amsfonts,amssymb]{article}
\def\erf{\mbox{\rm erf}}
\def\Erf{\mbox{\rm Erf}}
\newcommand{\extr}{\mathop{\rm extr}}
\begin{document}
\setcounter{page}{0}
\begin{titlepage}
\title{Optimally adapted multi--state neural networks trained with noise}
\author{R. Erichsen Jr. and W. K. Theumann\\
         Instituto de F\'{\i}sica, Universidade Federal do Rio Grande do
Sul,\\
  Caixa Postal 15051,\\
         91501--970 Porto Alegre, RS, Brazil }
\date{}
\maketitle
\thispagestyle{empty}
\begin{abstract}
\normalsize
\noindent The principle of adaptation in a noisy retrieval environment 
is extended here to a diluted attractor neural network of 
$Q$-state neurons trained with noisy data. The network is adapted to an
appropriate noisy training overlap and training activity which are
determined self-consistently by the optimized retrieval attractor overlap 
and activity. The optimized storage capacity and the corresponding 
{\it retriever} overlap are considerably enhanced by an adequate
threshold in the states. Explicit results for improved optimal performance
and new retriever phase diagrams are obtained for $Q=3$ and $Q=4$, with
coexisting phases over a wide range of thresholds. Most of the
interesting results are stable to replica-symmetry-breaking fluctuations.\\
\end{abstract}
\vspace*{0.5cm}
\newline \noindent
\end{titlepage}
\setcounter{page}{1}
\section*{1. Introduction}
\noindent Since the pioneering work of Hopfield \cite{Hop}, there has been much
interest in both the training and performance of attractor neural networks.
Training consists in encoding an appropriate synaptic matrix that enables
the network to store a macroscopic number of patterns, while the performance
of a network refers to the ability to retrieve one or a specific set of
stored patterns \cite{HKP}. Training and performance are usually 
thought of as separate stages in the operation of a network.

The retrieval performance of an attractor network can be studied 
in two different scenarios \cite{WS1}. One is characterized by a {\it fixed} 
synaptic prescription, as in the case of the Hopfield model \cite{Hop} or the 
maximally stable network (MSN) [4-6], while in the other one, the entire
space of synaptic interactions is searched for optimal performance whenever
there is a change in the retrieval environment. The synapsis in the first 
scenario are determined in an ordinary learning stage and the performance of 
the network is optimized separately in a given training environment. In the
second scenario one resorts to a continuously going on adaptive training 
process in which the network performance is optimized in an adiabatically
evolving retrieving environment \cite{WS1}. For each value of the noise 
parameter T (temperature of the retrieval dynamics), and storage ratio 
$\alpha$, the network has a unique interaction configuration, the so-called 
{\it retriever}.
This is in distinction to the retrieval performance that yields the phase
diagrams for the Hopfield model or the MSN, in which the interaction 
configuration determined in the separate learning stage is the same for all
T and $\alpha$ \cite{AEHW}.

Adaptive training processes seem to be biologically appealing as a mean 
to learn from the environment. The adaptive process in the second scenario
requires training the network with noisy patterns \cite{GSW,WS2} and
it is a procedure that does not separate the training process as a distinct
step from the operating stage of the network. The principle of adaptation
in a network of binary units consists in the search of the interaction
space for the optimized network performance adjusting the training noise
to be the same as the retrieval noise in each step of the adiabatically
evolving retrieving environment. Both noises refer to the Hamming distances
between the actual states of the network and the encoded patterns.

Training noises have been introduced in feedforward networks \cite{GT,Gy} in 
order to avoid overfitting
to training examples and in attractor networks with the purpose of enlarging
their basins of attraction \cite{GSW,WS2}. A slightly distorted set of
random patterns is presented to the network in the process of encoding the
synaptic matrix by means of a stepwise updating procedure following the
perceptron learning rule \cite{WS2}. The MSN is generated by an
infinitesimal amount of training noise and, except for low retrieval noise 
T and low load $\alpha$, the performance of the optimally adapted network
is clearly superior to that of the MSN \cite{WS1}. In particular, for low to 
moderate T and higher load $\alpha$, a second optimal solution in
interaction space appears for each value of the training noise in the
optimally adapted network. This solution is a weaker retriever which can
be interpreted as an attractor of self-adaptation.

The point is that the second retriever constitutes a further
solution to the optimization process, with its own interaction, in a
neighborhood of interaction space where there is no solution for the MSN.
This second, optimal solution, appears as a low performance solution 
in the absence or for low to moderate retrieval noise, with improving
performance, up to a certain point, as the retrieval noise is raised. Thus,
there has to be already a certain level of retrieval noise for the weak
retriever to have an interesting performance. Moreover, whenever the solution 
exists it is only within a narrow range of $\alpha$.

The principle of adaptation has been worked out, so far, only for
a network of binary neurons and the purpose of the present paper is to
explore the merits of an extension of the principle to a multi--state attractor
network in which both the neurons and the noisy training versions of the 
encoded 
patterns can be in $Q(>2)$ states. This adds two new dimensions to the study
of the performance of the network. First, the randomly distributed noisy 
patterns presented to the network in the training process introduce a training
activity $a_{t}$. Second, the firing rate of the neurons is determined by one
or more thresholds, or a growth parameter in the dynamical output function.
Thus, in the extension considered in this work, an evolving dynamical overlap
$m(\tau)$ and a dynamical activity $a(\tau)$ are generated at each time step 
$\tau$ of 
the neuron updating procedure. The search for the optimized network performance
by means of the extended adaptation principle consists now in the adjustment 
of the training overlap $m_{t}$ and the training activity $a_{t}$ to be 
$m_{t}=m(\tau)$ and $a_{t}=a(\tau)$, respectively, i.e., the same as the 
retrieval
overlap and dynamical activity in each step of the adiabatically  evolving 
retrieval environment. Adaptive performance in this wider sense is a
self--consistent procedure in which the retrieval environment continuously
optimizes the attractor performance of the network.

Networks of multi--state neurons have interesting features and applications.
Feedforward networks of such units can be used to study multi--class 
classification problems \cite{WRBM}, while multi--state attractor networks, 
which are useful for the recognition of various grey--toned patterns, are 
networks that have interesting inferential properties, by means of which the
storage capacity and the retrieval ability can be enhanced when they are 
trained with patterns of low activity [13--16]. Also, the categorization
ability can be improved in a multi--state network with hierarchical patterns. 
There has been lately considerable interest in such networks [17--19]. 

We consider an extremely diluted
network and, for simplicity, restrict ourselves to binary unbiased encoded 
patterns. The main emphasis of the paper is on the storage capacity, the 
quality of the performance of the strong and the second retrievers 
and on the characterization of the various phases that can appear. With that
purpose we produce explicit results for a network with $Q=3$ or $Q=4$
states. It will be shown that, within a finite range of a threshold parameter,
there is a considerable improvement of the storage capacity and in the high
performance of the second retriever solution, in the absence or for low 
retrieval noise, when compared with the optimally adapted network of binary 
neurons. In particular, we show that the second retriever may attain a fairly 
high retrieval overlap for small training noise in a regime where there is no
solution for the optimally adapted network of binary neurons. These are 
important results in the search for improvement of the behavior of attractor
neural networks. We restrict ourselves to finite--$Q$ state networks, in place 
of addressing the general (large--$Q$) case.

The outline of the paper is the following. In section 2 we extend the
training with noise procedure in the space of synaptic interactions to a
$Q$--state Ising network by means of a quenched optimization approach
\cite{WS1,WS3}, within the replica--symmetry Ansatz,
introducing a smooth cost function given by an average squared Hamming
distance.
The equations for the adaptation process in a
noisy retrieval environment are formulated in that section. The explicit 
results for the fixed--point behavior, the storage capacity and
the corresponding phase diagrams for self--adaptation for the three and the 
four--state models are discussed in section 3, and compared with the MSN. The 
domain of validity of the replica symmetric 
results is determined by the de~Almeida--Thouless lines \cite{AT}
in terms of the retrieval noise and the threshold in the dynamical updating
procedure. A summary and concluding remarks
are presented in section 4.

\section*{2. Training with noise and adaptation}
Consider a network of $N$ nodes with a dynamical variable $S_i(\tau)$, at
time step
$\tau$ on node $i$, that indicates the extent to which the unit on node $i$
fires. Each unit can be in any one of $Q$ Ising states
\begin{equation}
\sigma_k=-1+\frac{2(k-1)}{Q-1}
\label{2.1}
\end{equation}
in the interval $[-1,+1]$, for $k=1,\dots,Q$. A macroscopic set of $p$
binary
patterns $\{\xi_i^{\mu}=\pm1;\;\mu=1,\dots,p;\;
i=1,\dots,N\}$, with $p=\alpha C$, is encoded in the network in the
learning process, where $C$ is the connectivity of a node.
The patterns constitute a set of independent identically distributed random 
variables. Training consists in presenting to the network a noisy version
$\{R_i^{\mu}(\tau)\}$ of the   
patterns, at time $\tau$, and in the optimization of the network output after 
one time
step. This involves a dynamical process in the space of state configurations
of the network and, to keep the dynamics simple, we restrict ourselves to an
extremelly diluted network.
Each $R_i^{\mu}(\tau)$ is assumed to be in one of $Q$ states, $\sigma_k$, 
and can be thought of as an example of the pattern $\xi_i^{\mu}$.
Assuming that every
noisy pattern has the same overlap $m_t$ with the corresponding
pattern $\xi_i^{\mu}$, and that the activity $a_t$ is the same for all
patterns in the training set, we define
\begin{equation}
m_t=\frac{1}{N}\sum_i\xi_i^{\mu}\left\langle R_i^{\mu}(\tau)\right\rangle_R
\label{2.5}
\end{equation}
and
\begin{equation}
a_t=\frac{1}{N}\sum_i\left\langle
\left(R_i^{\mu}(\tau)\right)^2\right\rangle_R\;,
\label{2.6}
\end{equation}
where the brackets $\langle\dots\rangle_R$ denote averages over the
probability distribution of $R_i^{\mu}$. 
Thus, the noisy training inputs are constrained to satisfy the mean
$\langle R_i^{\mu}(\tau)\rangle_R=m_t\xi_i^{\mu}$
and variance
$\langle(R_i^{\mu}(\tau))^2\rangle_R-\langle R_i^{\mu}(\tau)\rangle_R^2=a_t-m_t^2$.

The normalized local field at node $i$,
due to the activity at the other nodes, is given by 
\begin{equation}
h_i(\tau)=\frac{1}{\sqrt{C}}\sum_{j=i_1}^{i_c}J_{ij}S_{j}(\tau)\;,
\label{2.2}
\end{equation}
where $J_{ij}$ is the synaptic connection between nodes $i$ and $j$,
independently in what state the dynamical variable $S_j$ is,
while $i_1,\dots,i_c$ denote the nodes
feeding node
$i$. The connections follow the spherical constraint $\sum_j
J_{ij}^2=C$, and
we consider the extremely diluted network in the limit of large 
connectivity in which $1\ll C\ll\ln N$. The one time--step dynamics is exact
in this limit.

We deal in this paper with the asymptotic, equilibrium configuration
$\{J_{ij}\}$, for the synaptic matrix elements of the learning process
that follows from a Langevin dynamics with a noise term. This involves an 
annealing temperature $T_{a}$ which takes care that the network does not get
trapped in local minima of the free energy. The distribution of equilibrium
states of the $J_{ij}$ can then be described by a canonical ensemble with
temperature $T_{a}$. 
Thus, there are two time scales in this approach: a short--time scale for the 
dynamical evolution of the synaptic matrix $\{J_{ij}\}$ and a long--time scale 
for the dynamical evolution of the training and of the retrieval parameters.

The dynamical variables are updated according to the rule
\begin{equation}
S_i(\tau+1)=g(h_i(\tau))\;,
\label{2.3}
\end{equation}
where $g(h_i(\tau))$ is the non--decreasing step function
\begin{equation}
g(x)=\sum_{k=1}^Q\sigma_k[\theta(b(\sigma_{k+1}+\sigma_k)-x)
		-\theta(b(\sigma_k+\sigma_{k-1})-x)]
 \label{2.4}
\end{equation}
shown in Figure~1 for $Q=3$ and $Q=4$ in which $\theta(x)$ is the unitary step 
function, $\sigma_0=-\infty$, 
$\sigma_{Q+1}=\infty$, $b\geq 0$ is the threshold parameter and $\sigma_k$
are the uniformly spaced Ising states of Eq.~(\ref{2.1}). According to
Eq.~(\ref{2.4}), there is a zero activity state whenever $Q$ is odd and none 
if $Q$ is even. 

For the adapted optimization a temperature $T$ is
introduced as a noise parameter, not to be confused with the annealing 
temperature $T_a$, to characterize the noisy retrieval environment. We assume 
a Gaussian thermal noise term added to the local field to write the one--step
output as
\begin{equation}
S_i(\tau+1)=g(h_i(\tau)+Tz)
\label{2.9}
\end{equation}
where $z$ has mean zero and unit variance.
The optimization, in the extremely diluted limit, consists in penalizing 
deviations
from the minimal output error in one time step on any node which is
independent
of the optimization on all the other nodes. Thus, it is sufficient to
consider
the cost function for a single node. We choose this to be
\begin{equation}
\label{2.10}
\sum_{\mu}d_i^{\mu}(\tau+1)\\
 =\sum_{\mu}\left\langle\left\langle\left[1-2\xi_i^{\mu}S_i(\tau+1)
  +S_i^2(\tau+1)\right]\right\rangle_z\right\rangle_R\;,
\end{equation}
where $d_i^{\mu}(\tau+1)$ is the average squared Hamming
distance to a stored pattern in which $\langle\dots\rangle_z$ denotes
the
average over the Gaussian thermal noise.
The training noise enters only through the local
fields, via
Eqs.(\ref{2.2}) and (\ref{2.9}). In the case of binary patterns, the 
local field is a Gaussian random variable with mean
%
$\langle h_i(\tau)\rangle_R=m_t\Lambda_i^{\mu}$
%
and variance
%
$\left\langle h_i^2(\tau)\right\rangle_R
 -\left\langle h_i(\tau)\right\rangle_R^2=a_t-m_t^2$,
%
in which
\begin{equation}
\Lambda_i^{\mu}=\frac{1}{\sqrt{C}}\sum_{j=i_1}^{i_c}J_{ij}\xi_j^{\mu}
\label{2.13}
\end{equation}
is the local field on node $i$ due to the pattern $\mu$.


The optimization of the Hamming distance between the one--step output of
the network in the noisy {\it training} environment and a given pattern in a 
network of binary
neurons is equivalent to finding the optimal output overlap after one
time step. In the case of a network of multi--state neurons, the Hamming
distance also depends on the activity through the local field, and our
first goal is
to find the optimal output Hamming distance $d(m_t,a_t)$, after one time
step, for a given training overlap {\it and} a given training activity. 

For that purpose, and for later use, we need the averages
\begin{equation}
S_{m_t,a_t}(\Lambda_i^{\mu})\equiv
\langle\langle S_i(\tau+1)\rangle_z\rangle_R
 =\frac{1}{2}\sum_{k=1}^Q\sigma_k \Erf(u_k,l_k;\Lambda_i^{\mu})
\label{2.14}
\end{equation}
and
\begin{equation}
S^2_{m_t,a_t}(\Lambda_i^{\mu})\equiv
\langle\langle S_i^2(\tau+1)\rangle_z\rangle_R
  =\frac{1}{2}\sum_{k=1}^Q\sigma^2_k \Erf(u_k,l_k;\Lambda_i^{\mu})
\label{2.15}
\end{equation}
which follow from Eq. (\ref{2.9}), where

\begin{equation}
\Erf(u,l,\Lambda)=\erf\left(\frac{u-m_t\Lambda}{\sqrt{2(a_t-m_t^2+T^2)}}\right)
  -\erf\left(\frac{l-m_t\Lambda}{\sqrt{2(a_t-m_t^2
  +T^2)}}\right)\;,
\label{2.16}
\end{equation}
with
\begin{equation}
u_k/2b=\sigma_k+1/(Q-1)\;, u_Q=\infty
\label{2.17}
\end{equation}
and
\begin{equation}
l_k/2b=\sigma_k-1/(Q-1)\;, l_1=-\infty\,.
\label{2.17a}
\end{equation} 

The quenched optimization approach \cite{WS1,WS3} requires the
introduction of the partition function
\begin{equation}
{\cal Z}(\beta)=\int\prod_j{\rm
d}J_{ij}\;\delta\left(\sum_jJ_{ij}^2-C\right)
 \exp\left[-\beta\sum_{\mu}d\left(\Lambda_i^{\mu}\right)\right]
\label{2.18}
\end{equation}
to obtain first an annealed average over the space of synaptic
connections
$J_{ij}$, in which $\beta=T_{a}^{-1}$ is the inverse annealing
temperature, and $d(\Lambda_i^{\mu})$ is the squared
Hamming distance, for a given configuration $\{\xi_i^{\mu}\}$ of 
encoded patterns, averaged over thermal and training noises. 
Its dependence on the noise parameters $m_t$, $a_t$ and $T$ is left implicit.
The quenched average free--energy is then obtained making use of the
replica method to write
\begin{equation}
\langle\ln{\cal Z}\rangle_{\xi}=\lim_{n\rightarrow 0}\frac{1}{n}\left(
 \langle{\cal Z}^n\rangle_{\xi}-1\right)\;,
\label{2.19}
\end{equation}
where $\langle\dots\rangle_{\xi}$ denotes the average over the set of
stored patterns
$\{\xi_i^{\mu}\}$. Using the standard technique in the space of synaptic
interactions, with the assumption of replica symmetry
\cite{Ga2,Ga3}, we obtain the optimal one--step output Hamming distance for
training
\begin{eqnarray}
d(m_t,a_t)&=&-\lim_{\beta\rightarrow\infty}\frac{1}{\alpha\beta C}
 \langle\ln{\cal Z}\rangle_{\xi}\\
 &=&\extr_x\left\{\int{\rm D}y\;\min_{\lambda}\;
 F(\lambda,x,y)-\frac{1}{2\alpha x}\right\}\nonumber
\label{2.20}
\end{eqnarray}
as a function of the overlap $m_t$ and activity $a_t$ of the noisy input
patterns, in which ${\rm D}y={\rm e}^{-y^2/2}{\rm d}y/\sqrt{2\pi}$
is a Gaussian measure and
\begin{equation}
F(\lambda,x,y)=d(\lambda)+\frac{(\lambda-y)^2}{2x}\;.
\label{2.22}
\end{equation}
Here, $d(\lambda)$ is the squared Hamming distance averaged over
$\xi_i^{\mu}$, while $x=\beta(1-q)$ and
\begin{equation}
q=\frac{1}{C}\sum_jJ_{ij}^{\rho}J_{ij}^{\sigma}
\label{2.23}
\end{equation}
for all $\rho\ne\sigma$ is the spin--glass order parameter for the problem.

The optimization in the training process amounts to take the limits
$\beta\rightarrow\infty$ and $q\rightarrow 1$ keeping $x$ finite. A single
solution in the space of interactions is thus obtained out of the full
multiplicity of solutions when $q\rightarrow 1$ \cite{Ga3}. The
minimization with respect to $\lambda$ yields
\begin{equation}
y(\lambda)=\lambda+x d'(\lambda)\;,
\label{2.24}
\end{equation}
where $\lambda=\lambda(y)$ is the inverse function of $y(\lambda)$. On the
other hand, the extremum in $x$ gives the saddle--point equation
\begin{equation}
\alpha^{-1}=\int{\rm D}y\left[\lambda(y)-y\right]^2\;,
\label{2.25}
\end{equation}
which determines the storage capacity $\alpha$ for a given training 
environment.

In cases where $\lambda(y)$ is a multivalued function of $y$, which is
the case for $Q\geq 2$, there may be one or more transitions, each with a 
fixed $y_0$ between an upper and a lower value $\lambda_>$ and $\lambda_<$, 
respectively, ruled by a Maxwell construction
\begin{equation}
\int_{\lambda_<}^{\lambda_>}{\rm d}\lambda\;y(\lambda)=y_0(\lambda_>
 -\lambda_<)
\label{2.31}
\end{equation}
where $y_0=y(\lambda_<)=y(\lambda_>)$. It turns out that the function
$F(\Lambda,x,y)$ is the same on both sides of the ``first--order''
transition.

The optimal output Hamming distance for training becomes then
\begin{equation}
d(m_t,a_t)=\int{\rm D}y\;d(\lambda(y))\;.
\label{2.26}
\end{equation}
It is convenient to introduce the distribution of the local fields due to
the encoded patterns, defined as [3--5]
\begin{equation}
\rho(\Lambda)=\left\langle\left\langle\delta\left(\Lambda-\frac{1}{\sqrt{C}}
\sum_jJ_{ij}\xi_j\right)\right\rangle_J\right\rangle_{\xi}
\label{2.27}
\end{equation}
where the ensemble average $\langle\dots\rangle_J$ is performed with the
partition function ${\cal Z}(\beta)$, Eq.~(\ref{2.18}). It turns out that
this distribution becomes
\begin{equation}
\rho(\Lambda)=\int{\rm D}y\;\delta(\Lambda
 -\lambda(y))\,,
\label{2.28}
\end{equation}
and the transition between the lower and upper bonds, $\lambda_<$ and 
$\lambda_>$ respectively, implies a gap in the distribution of local fields
$\rho(\lambda)$ whenever $\lambda(y)$ is a multivalued function of $y$.

The optimal one--step output Hamming distance for training with noise may now 
be written as
\begin{equation}
d(\tau+1)=1-2f_{m_t,a_t}(m_t,a_t)+g_{m_t,a_t}(m_t,a_t)
\label{2.36}
\end{equation}
where
\begin{equation}
f_{m_t,a_t}(m,a)=\int{\rm d}\Lambda\;\rho_{m_t,a_t}(\Lambda)
  S_{m,a}(\Lambda)
\label{2.37}
\end{equation}
is the optimized overlap between the encoded patterns and their noisy versions
and
\begin{equation}
g_{m_t,a_t}(m,a)=\int{\rm d}\Lambda\;\rho_{m_t,a_t}(\Lambda)
 S_{m,a}^2(\Lambda)\;.
\label{2.38}
\end{equation}
is their optimized activity. The distribution of the local fields,
$\rho_{m_t,a_t}(\Lambda)$, is a characteristic property of the training
set and, as such, it depends on $m_t$ and $a_t$. 

The formal results presented so far assume that replica symmetry holds in the
space of interactions. The condition for local stability of the replica
symmetric saddle--point can be writen as \cite{WS3,Bo}
\begin{equation}
\alpha^{-1}>\int{\rm D}y\;\left[\lambda'(y)-1\right]^2\;,
\label{2.39}
\end{equation}
in which $\lambda'={\rm d}\lambda/{\rm d}y$,
and this is to be solved together with Eq.~(\ref{2.25}).
When the distribution of the local fields has a gap, $\lambda'$ diverges and
the condition cannot be satisfied. Then, the network becomes unstable to
replica--symmetry--breaking
fluctuations. The limiting load for which Eq.~(\ref{2.39}) is still
satisfied
yields the de Almeida--Thouless (AT) line, $\alpha_{AT}(T)$ \cite{AT}.
The dependence on the retrieval noise
$T$ comes from $\lambda$. Note that the AT line must lie
within the one--band region or, at most, on the band--merging surface
where the gap in
$\rho(\lambda)$ disappears \cite{Bo}. This completes the formal description of
the training process in itself. In order to become optimally adapted, we 
consider now the retriever process.

The calculation of the one--step output Hamming distance between {\it any} 
input state $\{S_i(\tau)\}$ and a given encoded pattern in a noisy retrieval 
environment, with temperature $T$, is now obtained as follows. First, the 
training parameters $m_t$ and $a_t$ in Eqs.~(\ref{2.14})--(\ref{2.16}) are 
replaced by the overlap $m(\tau)$ and the dynamical activity $a(\tau)$ of the 
noisy {\it retrieval} state $\{S_i(\tau)\}$, expressed respectively as 
Eqs.~(\ref{2.5}) and (\ref{2.6}) with $\{S_i(\tau)\}$ 
in place of the the noisy pattern $\{R_i^{\mu}(\tau)\}$.
The one--step output Hamming distance in the {\it retrieval} 
environment is now given by an expression similar to Eq.~(\ref{2.36}), 
depending on the pair ($m_t,a_t$) 
through the distribution of local fields {\it and} on the pair ($m,a$) through 
the present state of the network as given, literally, by Eqs.~(\ref{2.37})
and (\ref{2.38}). 
%
%

Now, the training overlap $m_t$ and the training activity $a_t$ which give the
optimal performance for retrieval at a fixed temperature $T$, storage level
$\alpha$ and threshold parameter $b$, are given by the adaptation principle. 
The optimal adaptation consists in a search in the space of interactions
$\{J_{ij}\}$ simultaneously with a search in the space of state configurations
$\{S_i(\tau)\}$. The best adapted performance of the network is attained by 
adjusting the training noise and activity to the same level as the retrieval 
noise and activity. For the parallel dynamics  in the extremely diluted 
network we are dealing with, the stable fixed point of the set of equations
\begin{equation}
    f_{m,a}(m,a)=m
\label{3.6}
\end{equation}
and
\begin{equation}
    g_{m,a}(m,a)=a
\label{3.6a}
\end{equation}
gives at the same time the optimal training condition and the optimized
performance. The stable fixed point for each value of the synaptic noise 
parameter $T$, the storage ratio
$\alpha$ {\it and} the threshold parameter $b$ is a {\it retriever}, for
which the network has a unique interaction configuration. In 
other words, in distinction to the usual phase diagrams for retrieval, every 
point of the phase diagrams that will be discussed next represents a different 
network.

\section*{3. Results and discussion}
We present next the results for the optimally adapted retrievers.
The rich structure of locally stable states and the corresponding phase
diagrams for self--adaptation that arise as the threshold parameter
$b$ is increased will be discussed now, separately for $Q=3$ and $Q=4$.

\subsection*{3.1. Three--state network}
To illustrate the role of the threshold parameter $b$, we discuss first
the fixed--point solutions for $m$ and $a$ and the corresponding phase
diagram for $\alpha$ vs. $b$, in the absence of retrieval noise shown in
Figure~2. For fixed $b$ within
the range $0\leq b\leq 0.57$ and $0\leq\alpha\leq\alpha_1(b)$, there is a
perfect retriever with $m=1=a$ which is the only stable fixed point, and a
solution with $m=0$ and either $a\ne 0$ or $a=0$, which is an unstable fixed
point. This suggests that one can conceive a network capable of perfect
retrieval operating with a limited threshold, as long as the training is
with infinitesimal noise $m_t=1^-$ and almost full activity $a_t=1^-$. The
corresponding retriever is that of the MSN. 

The line $\alpha_1(b)$ deserves further attention. It is the upper
bound of the region where the perfect 
retriever is the only attractor in the retriever dynamics with a wide basin
of attraction for self--adaptation. Beyond that line, the basin of attraction of this retriever is 
greatly reduced in the three--state network, as will be seen next.
Thus, for increasing $b$,
in the small $b$ regime, there is an enhancement of the associativity of the 
network, as long as $\alpha_1$ is an increasing function of $b$.

A new pair of stable and unstable
fixed points appears discontinuously at $\alpha_1(b)$.
The stable fixed point represents a new retriever of weaker attractor
overlap and reduced activity. Note, however, that for low to moderate $b$
(illustrated in the inset by $b=0.5$), there is a considerably enhanced
retrieval overlap when compared with the overlap for the optimally adapted 
network of binary units \cite{WS1}. The second retriever has a rather wide
basin of attraction for this larger retriever overlap.
This higher performance can be attained through training with low--noise 
patterns with moderately high activity. For the threshold $b\simeq 0.3$
that maximizes $\alpha_1(b)$, the improvement in storage capacity with the 
{\it same} retrieval overlap as that of the network of binary neurons is about 
$20\,\%$. However, as one would expect,
the performance deteriorates with a further increase in the threshold $b$.

The second stable fixed point means that there exists a second training
condition, with higher noise, which results in a network with lower, but
still optimal performance when compared with other three--state networks in its
vicinity of the space of interactions, for this training condition. The 
unstable fixed--points are repelors of the self--adaptation dynamics 
\cite{WS1}.

The overlap of this second retriever vanishes continuously as
$\alpha$ increases approaching $\alpha_2(b)$. For
$\alpha_2(b)\leq\alpha\leq\alpha_c(b)$, the perfect retriever and a
non--retriever with $m=0$, and either a finite or no activity, are the only 
stable fixed--point solutions. The non--retriever state with $a\ne 0$ appears 
as a self--sustained activity phase, which has been discussed first for a 
diluted network with a
Hebbian learning rule \cite{Ye}. When the activity is zero the network
stops operating.

The presence of a non--retriever with finite activity follows from the
fixed--point solution for $(m,a)$ when $m=0$ is a stable fixed--point. The
expression for $S_{m,a}^2(\Lambda)$ becomes then independent of the local field
$\Lambda$ and, hence, of $x$ and $\alpha$. The fixed--point values for
$a$ are then given by the solutions of the equation
%
$
a=1-\erf(b/\sqrt{2(a+T^2)})
$.
%
The solution $a=0$ is stable for all $b$, when $T=0$. There is a second
stable fixed point that decreases monotonically from $a=1$, at $b=0$, and
disappears discontinuously at $b\simeq 0.57$ when the value $a\simeq 0.23$ is
reached. This is the origin of the ``tricritical'' point in the phase diagram 
for $\alpha$ vs. $b$, where the line of continuous transitions for the overlap
becomes discontinuous. We come back to this point below. It is important to
remark that the term ``transition'' here only means that the network changes
from one retriever state to another one. We remind that it is not meant as an 
usual thermodynamic phase transition, since each point of the phase diagram
corresponds to a different network.

Finally, when $\alpha$ reaches the critical storage capacity
$\alpha_c(b)$, given by
\begin{equation}
\alpha_c^{-1}(b)=\int_{-\infty}^b {\rm D}y(b-y)^2
\label{4.2}
\end{equation}
the perfect retriever is destabilized.

Consider next the case where $0.57<b\leq 0.82$. For
$0\leq\alpha<\alpha_1(b)$,
there is again a perfect retriever which is a stable fixed--point solution. In
addition, a pair of stable and unstable fixed points appears. The stable
fixed point is a non--retriever with $m=0$ and either $a\ne 0$ or $a=0$.
A new pair of stable and unstable fixed points appears discontinuously at
$\alpha_1(b)$. The stable fixed point is, again, a retriever of weaker
attractor overlap
and reduced activity. However, as $\alpha$ approaches $\alpha_2(b)$, this 
second retriever vanishes {\it discontinuously} and, thus, there is a
changeover from the line of continuous transitions $\alpha_2(b)$ when $b$
increases and reaches a tricritical point at $b\simeq 0.57$. 
When $\alpha$ increases beyond $\alpha_2(b)$,
the perfect retriever and the non--retriever are, again, stable fixed--point
solutions, and the perfect retriever, which has a narrow basin of attraction, 
is destabilized when the critical
$\alpha_c(b)$ is reached. When $b$ is increased, the retriever of weaker 
attractor overlap disappears at
$b\simeq 0.82$, and beyond this point the perfect retriever is the only stable
fixed point with finite overlap for $0\leq\alpha\leq\alpha_c(b)$.

Now we discuss the stability of the replica symmetric solution. First, 
the strong retriever state is always stable to
replica--symmetry--breaking fluctuations below $\alpha_c$. Thus, at most the 
weak retriever can become unstable. In view of this, we mapped out the region
of the phase diagram where the stability condition, Eq.~(\ref{2.39}), is not 
satisfied for the
weak retriever state, and this is shown as the shaded area in Figure~2, the
dash--dotted line being the AT line. Furthermore, we found that this line
corresponds to the appearing of a gap in the distribution of local fields.

The phase diagram also yields the optimal basin boundary of the 
self--adaptation dynamics for a given $\alpha$. Thus, as $\alpha$ increases 
for $b\leq 0.82$ the strong retriever is a ``wide'' retriever for 
$\alpha<\alpha_1(b)$, since it is
the only attractor in the self--adaptation dynamics. For $\alpha>\alpha_1(b)$,
the strong retriever becomes a ``narrow'' retriever which coexists with the
weak retriever. Finally, for $b>0.82$, the strong retriever is a narrow
retriever that coexists with the non--retriever state for all
$\alpha\leq\alpha_c(b)$.

We consider next the results in the presence of retrieval noise $T$.
In the case of a small to moderate threshold where the strong and weak
retriever coexist, say, for $b=0.5$, the phase diagram for $T$ vs. $\alpha$
is not very different from the phase diagram for the network of binary
units. The strong and the weak retriever coexist now over a wider range of
$\alpha$ but the strong retriever disappears, as one would expect, for a
lower $T$. 
More interesting are the results for the phase diagram and the underlying
fixed--point
solutions for the overlap and the activity when $b=1$, shown in Figure~3.
This threshold is typical of an optimally adapted network that has a
perfect retriever as the only stable fixed point with non--zero overlap at
$T=0$. For fixed and low 
$T\leq 0.4$, there is a strong retriever with rapidly
decreasing $m$ and $a$ when $\alpha$ comes close to the line $\alpha_c(T)$,
where both parameters vanish discontinuously. There is a second stable
fixed point with $m=0$ and $a\sim 0$, for all $\alpha\geq 0$, and this
non--retriever is the only stable solution for $\alpha>\alpha_c(T)$. There is 
also an unstable fixed
point for $m$ and $a$ throughout the range $0\leq\alpha\leq\alpha_c(T)$
that separates the basin of attraction for self--adaptation of the two stable 
fixed points, and
indicates that the strong retriever is a narrow retriever in this
interval.

An increase in retrieval noise can be of use for the enhancement
of the performance of the single, strong retriever, with a moderately large
threshold, as in the present case of $b=1$. Indeed,
for $0.4\leq T\leq 0.8$, the non--retriever becomes an {\it un}stable
fixed point for $\alpha$ below the line $\alpha_1(T)$, leaving the
strong retriever as a wide retriever.
The overlap and the activity change discontinuously as $\alpha$ goes
through $\alpha_1(T)$. For $T\geq 0.8$, the overlap of the wide retriever
vanishes continuously as $\alpha$ approaches $\alpha_c(T)$. The results
shown here confirm the general expectation that one cannot attain the best
retriever overlap (as we have here for the narrow retriever) together with the
best associativity, as for the wide retriever, in the same network except at
the phase boundary.

The AT line coinciding with the locus where the gap closes down is also shown
in Figure~3, and the region to the right of the line up to the $\alpha_c$
line is stable to
replica--symmetry--breaking fluctuations. Thus, it seems that the part of the
discontinuous transition line $\alpha_c(T)$ that is close to the
tricritical point where the changeover to the line of continuous transitions 
takes place, is marginally stable. We also argue
that for low $T$ the line $\alpha_c(T)$ may be almost correct, since
$\alpha_c(0)$ is the critical capacity of
the MSN, which corresponds to a stable point. Note that the line $\alpha_c(T)$
of discontinuous transitions has an upper part of infinite slope which
should also be correct since one would not expect a reentrant behavior for
$\alpha_c(T)$. Finally, for comparision, we also show the phase boundaries for
the MSN and conclude that the optimally adapted network with three--state
neurons has an improved performance in the presence of retrieval
noise.

\subsection*{3.2. Four--state network}
To see now the effects of the threshold in the optimally adapted four--state
network, we present first the results for the fixed--point solutions for
the overlap and the activity in Figure~4. Depending on the value of $b$
there may be
a domain in the values of $\alpha$ in which there are up to three stable
fixed--point solutions with non--zero $m$, one for a perfect retriever
and the other ones for weaker retrievers.
The perfect retriever exists up to a critical $\alpha_c(b)$, given by
\begin{equation}
\alpha_c^{-1}(b)=\int_{-\infty}^{4b/3} {\rm D}y\left(\frac{4b}{3}-y\right)^2\;.
\label{4.3}
\end{equation}

It turns out that there is a load $\alpha_1(b)$ for all $b$, where a weak
retriever, which may or may not be the only one, appears discontinuously as
$\alpha$ attains that point. For $b\leq 0.65$, it is the only weak retriever,
as can be seen in the phase diagram for $\alpha$ vs. $b$ shown in
Figure~5. Note that, also for the four state network, $\alpha_1$ increases
with $b$ in the small $b$ regime with a considerable enhancement of the strong
retriever as a wide retriever.
The perfect and the weak retriever coexist with increasing 
$\alpha$ until either the weak retriever disappears continuously at 
$\alpha_2(b)$, which is the case
for $b\leq 0.44$, or the strong retriever ends at $\alpha_c(b)$ for
$0.44\leq b\leq 0.65$. In the latter case, the weak retriever of
non--zero overlap remains as the only attractor of self--adaptation up to
$\alpha_2(b)>\alpha_c(b)$.

On the other hand, for $b$ well above $0.65$, a second weak retriever
(${\rm WR}_2$) appears discontinuously as $\alpha$ attains the line 
$\alpha_4(b)$ while the
first weak retriever (${\rm WR}_1$) extends up to a quite higher load
$\alpha_3(b)$, where the state of the network changes discontinuously to
the non--retriever
state. The overlap of the ${\rm WR}_2$ vanishes continuously as $\alpha$
approaches $\alpha_2(b)$. The two weak retrievers coexist for
$\alpha_4(b)\leq\alpha\leq\alpha_2(b)$. Note that both the line where
the first weak retriever disappears and the domain of $\alpha$ where the second
weak retriever exists may lie well above the critical capacity $\alpha_c(b)$
for the existence of the perfect retriever.

The situation can become more involved for intermediate values of $b$,
shown by the inset in Figure~5. Around the endpoint C of the wedge of
discontinuous transitions lines, the second weak retriever can be reached 
continuously from the first one.

It is interesting to note that, for large $b$, the ${\rm WR}_1$ state has an
asymptotic overlap and activity $m\sim 1/3$ and $a\sim 1/9$, respectively.
These correspond to the storage of binary patterns in a network with only the
microscopic states $S_i=\pm 1/3$ being activated. These are, practically, the
only states favoured in the high--$b$ regime, since the states $S_i=\pm 1$ can
only become active by means of high local fields which are extremely unlikely
in the absence of retrieval noise. Indeed, we found that the line $\alpha_3(b)$
goes to the critical value $\alpha_c=2$ for the optimal network of binary 
units with increasingly large $b$. Thus, as expected, the behavior of the 
network in the large--$b$ limit should become that of the MSN with reduced 
overlap and activity.

%
%

The phase diagram in Figure~5 also provides the optimal basin boundary
of attraction, for a given $\alpha$ and $b$. For $b=1$, say, the strong
retriever is a wide retriever for $\alpha<\alpha_1(b)$, and a
narrow retriever when $\alpha_1(b)\leq\alpha\leq\alpha_c(b)$. On the other 
hand, in the interval $\alpha_c(b)\leq\alpha\leq\alpha_4(b)$,
the weak attractor with higher overlap is a wide retriever, since it
is the only attractor for the self--adapting dynamics in this interval.
In distinction, in the interval $\alpha_4(b)\leq\alpha\leq\alpha_3(b)$
that weak retriever is a narrow retriever, that coexists with ${\rm
WR}_2$ if $\alpha\leq\alpha_2(b)$ and with the non--retriever state otherwise.

To discuss the validity of the replica symmetric results note that, whenever
two weak retrievers coexist in the phase diagram, each one has to be analyzed
separately since they refer to different levels of training noise, such that
one may correspond to a gapless local field distribution and the other may not.
The AT line is the dash--dotted line shown in Figure~5, that
starts on the boundary $\alpha_1(b)$ where the single weak retriever appears
for small $b$ and it merges with $\alpha_4(b)$ around $b=0.8$. That retriever 
is stable to replica--symmetry--breaking fluctuations above the AT line. The
${\rm WR}_2$ is unstable around C and is stable in the strip
$\alpha_4(b)\leq\alpha\leq\alpha_2(b)$, whereas the ${\rm WR}_1$ is unstable
everywhere below and at the boundary $\alpha_3(b)$.
The left part of the boundary $\alpha_1(b)$ is marginally stable, as well as 
the boundary $\alpha_c(b)$ for the perfect retriever.


\section*{4. Summary and concluding remarks}
The principle of adaptation, formulated earlier for a network of binary 
neurons, has been extended in this work to study the training and performance 
of optimally
adapted attractor neural networks of multi--state neurons trained with noisy
inputs in the presence of a noisy retrieval environment. Explicit results where
obtained for the optimal attractor overlap and the optimal dynamical activity
as functions of the retrieval noise $T$, the load $\alpha$ and the threshold
$b$, for a network with dilute connectivity. The maximum storage capacity was
also obtained as a function of $b$ and $T$ and explicit retriever phase 
diagrams of performance and associativity of the retrievers are exhibited for 
a network of three or four--state neurons. These are phase diagrams for
{\it self--adaptation}, in distinction to phase diagrams for attraction, as
pointed out in ref.~\cite{WS1}. We remind the reader that, as 
pointed out by Wong and Sherrington, coexisting retrievers are solutions for 
different networks, which should correspond to distinct synaptic interactions.

An important issue of this work concerns the improvement in the 
associativity of multi--state networks, when the width $b$ of the
intermediate states increases, in the small $b$ regime. The enhanced 
performance of the second retrievers has also been emphasized.
This is important because they are optimal {\it retriever} solutions on their 
own, rather than weaker {\it retrieval} solutions  for the optimal 
network configuration, if such solutions exist \cite{WS1}. 
We have shown that an improvement of
the performance of the second retriever in the optimally adapted network 
with multi--state units can be attained with relatively small training 
noise and large--activity input patterns. In practical terms, this may 
be a more accessible situation than training with an infinitesimal amount of
noise and almost full activity. Furthermore, we have shown that the storage 
capacity
of the second retriever is a non--monotonic function of the threshold $b$ 
with an increasing capacity for small $b$. With a moderately large threshold,
as in the case of $b=1$ for the three--state network, an increase in retrieval
noise $T$ may help to enlarge the basin of attraction of the single, strong
retriever. This can be understood noting that the increase in the noise should
aid to overcome the large gap in the local field in firing the units when the 
network has been trained
with a moderate training noise. These are important results in the 
search for improvement of the behaviour of attractor neural networks.

The work presented here is restricted, for simplicity, to binary encoded 
patterns. On the basis of results we obtained for three or four--state 
patterns, we argue that this should not be a serious restriction. What is 
important is that the states of the noisy training set $\{R_i^{\mu}(\tau)\}$ 
have the same degrees of freedom as the arbitrary input set $\{S_i(\tau)\}$ for
retrieval. This requires the introduction of a training activity $a_t$ 
in the noisy inputs, in order to optimize both the training and the 
adaptation process in the $Q$--state network.

We have found, in accordance with earlier works, that networks are
specialized \cite{WS1,AEHW}.
Indeed, one cannot attain the best storage capacity for all $T$ and $b$ in a
single network. Even if $b$ is fixed the storage capacity of the strong
retriever will be that of the
MSN only at very low $T$ and it will become that of the Hopfield model at high
$T$.

All the results were obtained with the assumption
of replica symmetry in the space of synaptic interactions and the limit of
validity of this assumption has been established finding the 
de~Almeida--Thouless
lines $\alpha_{AT}(b)$ at $T=0$ and $\alpha_{AT}(T)$ for a given $b$. These
lines coincide with the band--merging lines for the distribution of the local
field. Due to the presence of optimal solutions for small--to--moderate
training noise, there are gaps in the distribution of the local fields over
sizeable domains of the phase diagram which are not stable to
replica--symmetry--breaking fluctuations. Nevertheless, interesting phase 
boundaries and domains of the phase diagrams are
stable or, at worst, marginally stable, confirming the validity of our results.
Indeed, the enhancement of 
the line $\alpha_1(b)$, where the second retriever appears for small training 
noise and large activity, both for $Q=3$ and $Q=4$, lies on the replica 
symmetric side of the AT line. Furthermore, the interesting weak retriever 
lies completely on this side. That is also the case for the tricritical point
and the first--order transition line, $\alpha_2(b)$, for the three--state
network, which at worst becomes marginally stable. Furthermore, the phase 
diagram for $T=T(\alpha)$ reveals that the line $\alpha_2$ of continuous
transitions is stable to replica--symmetry--breaking fluctuations, for both 
$Q=3$ and $Q=4$ and all $b$. In view of these results, it does not seem
worthwhile to pursue a calculation beyond the replica--symmetry Ansatz.

A closer look at our results reveals that although the critical capacity
$\alpha_c$, where the strong retriever terminates, decreases faster with
increasing $b$ for the four--state than for the three--state network, the 
trend is opposite for the lower and upper critical storage ratio $\alpha_1$
and $\alpha_2$ respectively, for the presence of a second retriever in the 
low--$b$ regime. This suggests that the role of the threshold could become 
even more important in optimally adapted higher Q--state networks. The 
extended 
principle of adaptation of the present work assumes that both, the training 
overlap and the training activity become continuously adapted to
the noisy retrieval environment. In particular, the training activity 
follows the changes in the dynamical activity characteristic of the $Q$ states 
of the units, and this makes difficult the study of the optimally adapted 
network for general $Q$. It may be possible to study a weaker version of the 
extended adaptation principle for the graded response network in which the 
training activity remains fixed. This, and other questions, will be considered 
in future work.

\section*{Acknowledgements}
We thank J. F. Fontanari and D. Boll\'e for critical comments,
and one of us (WKT) thanks the kind hospitality of the Institute for
Theoretical Physics of the Catholic University of Leuven, Belgium, where
part of the work was written. The research of one of us (WKT) was
supported by CNPq (Conselho Nacional de Desenvolvimento Cient\'{\i}fico e
Tecnol\'ogico, Brazil), and the work was supported in part by FINEP
(Financiadora de Estudos e Projetos, Brazil). A grant from CNPq on a
neural network project is gratefully acknowledged.

\newpage


%
%
\newpage
\section*{Figure captions}
\noindent
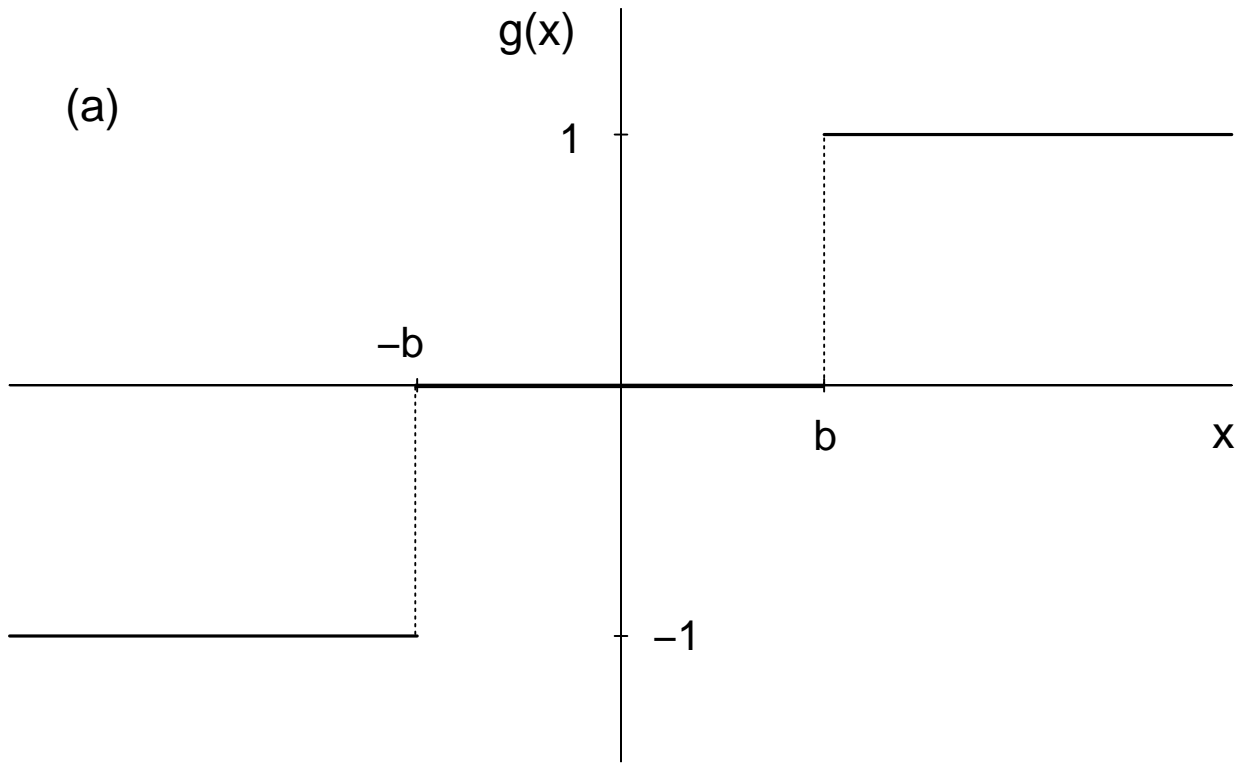
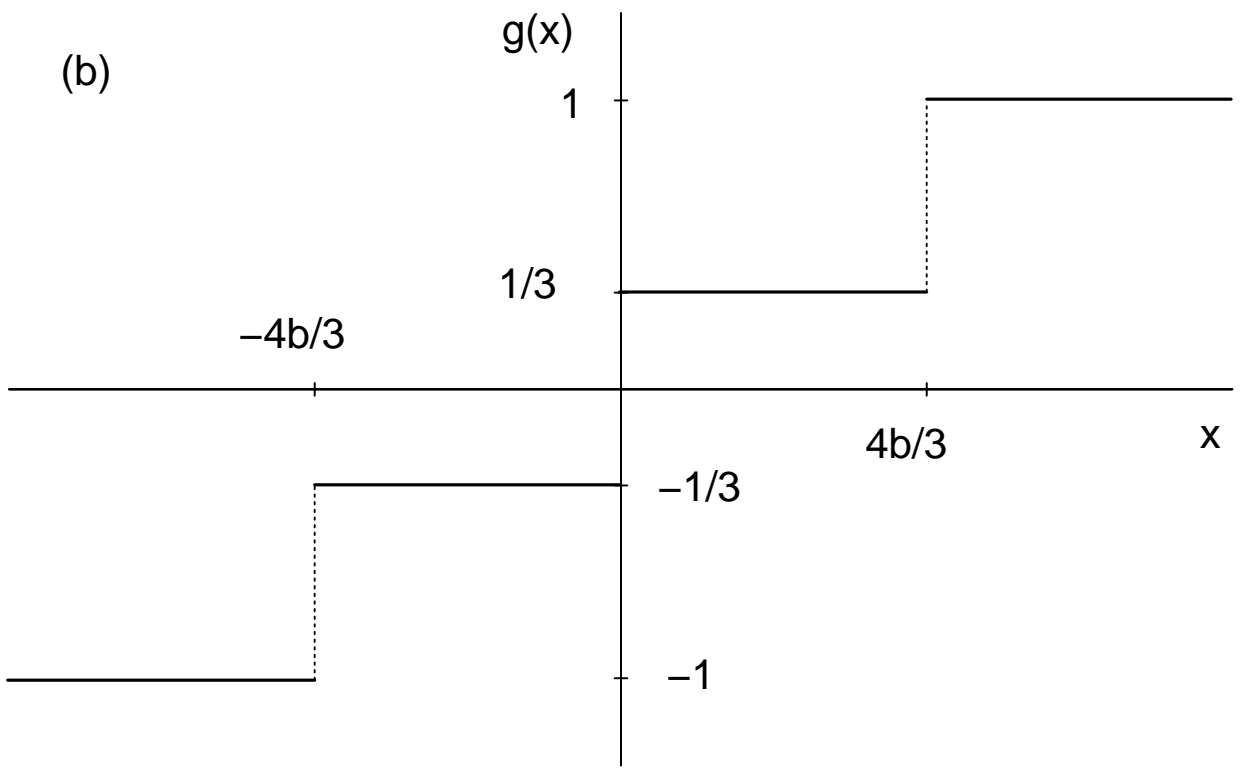
{\bf Figure 1:} The non--decreasing step function $g(x)$ for 
$Q=3$ (a) and $Q=4$ (b).
\\
\noindent
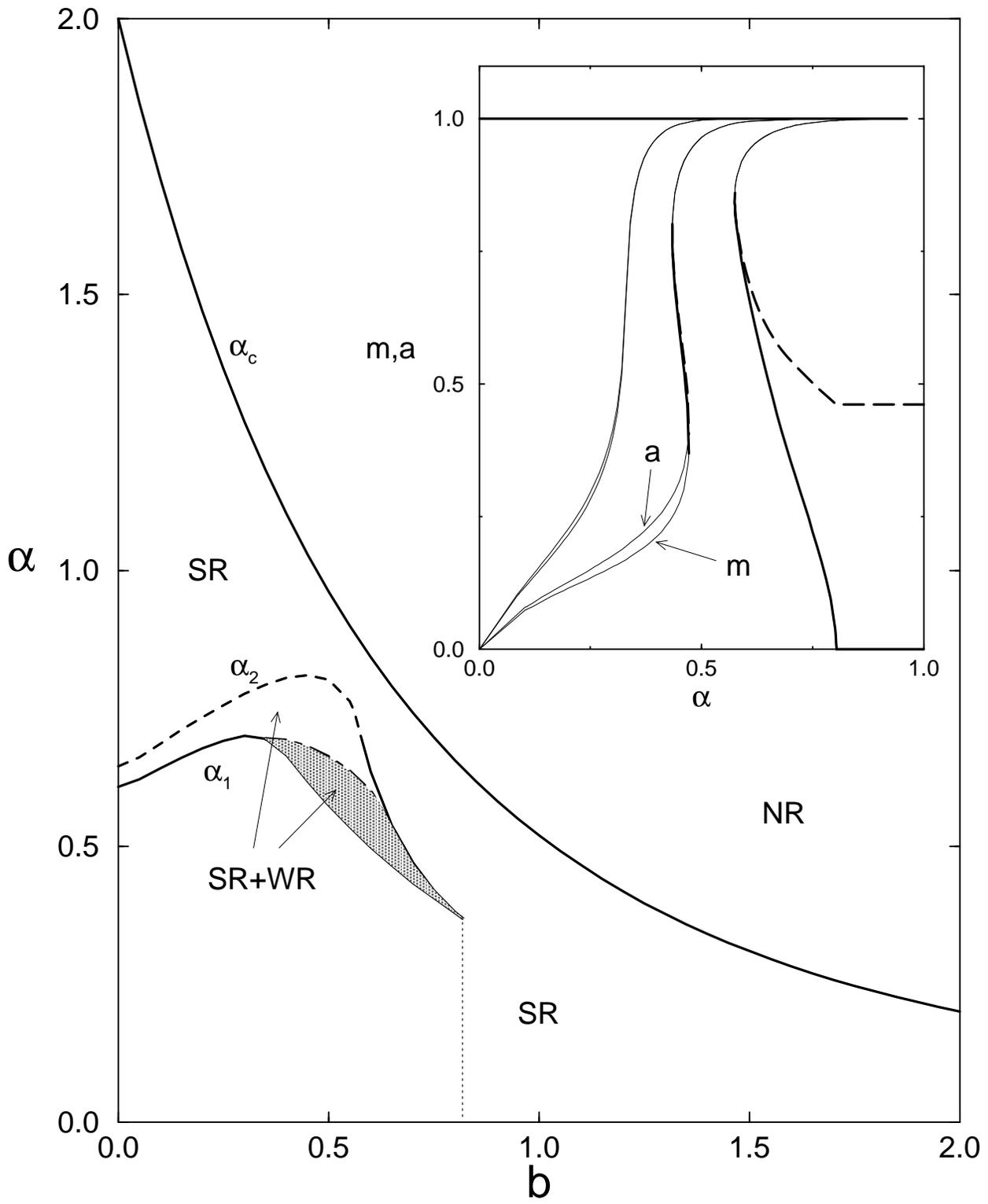
{\bf Figure 2:} Phase diagram for the load $\alpha$ as a function of the
threshold $b$ for $Q=3$, at $T=0$, and the corresponding optimal overlap
$m$ (solid lines) and activity $a$ (dashed lines) for $b=0.5$ (right),
$b=0.7$ (center) and $b=0.9$ (left), in the inset. Unstable fixed--point
solutions are shown in light lines. SR and WR are strong and weak 
retrievers, respectively. The SR is a wide retriever at the left of the light 
dotted line and below $\alpha_1(b)$.
Solid lines in the phase diagram indicate discontinuous 
transitions and a dashed line a continuous transition. The dash--dotted line 
is the de~Almeida--Thouless line (cf. the text). The WR is unstable to 
replica--symmetry--breaking in the shaded area.
\\
\noindent
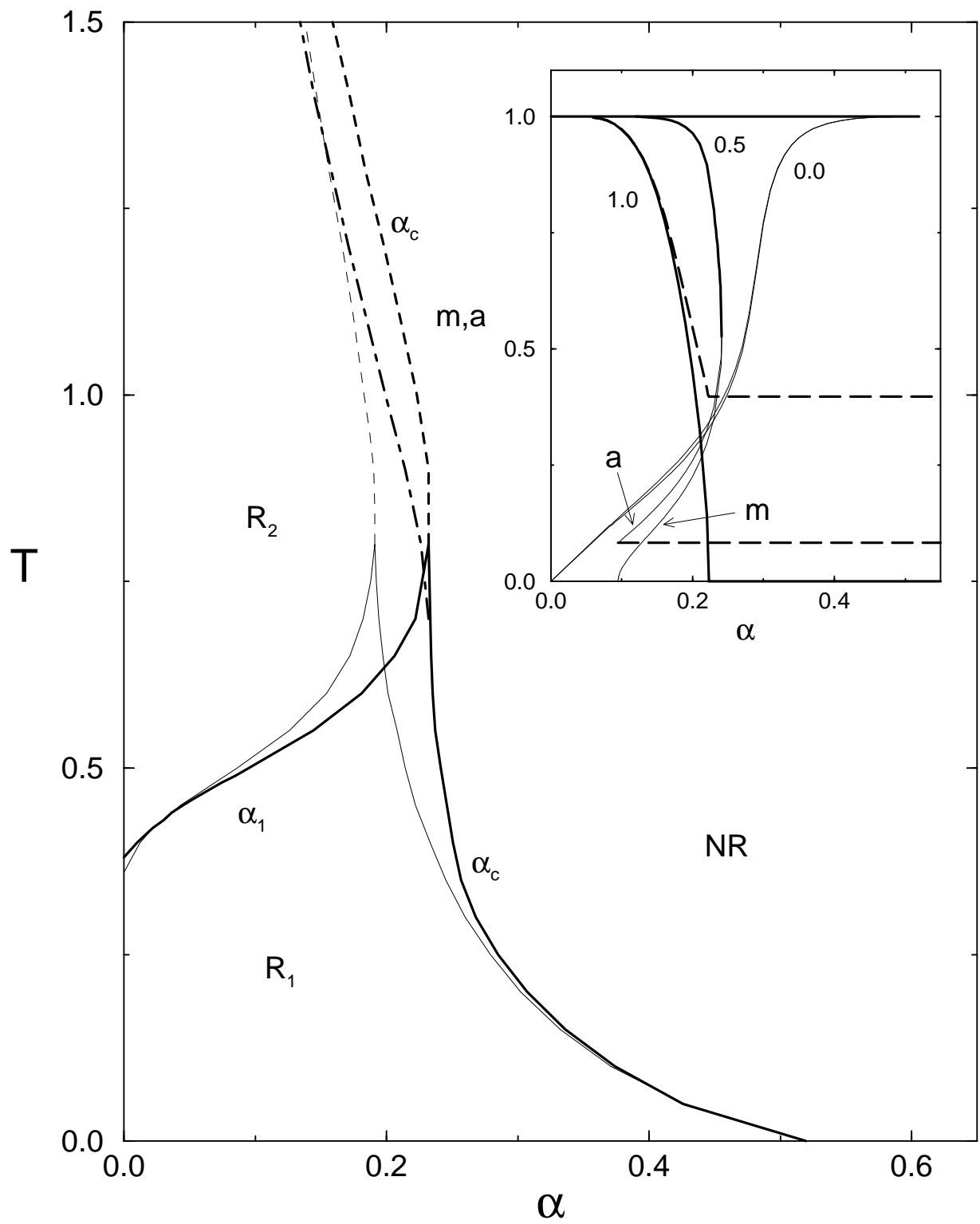
{\bf Figure 3:} Phase diagram, for $T$ vs. $\alpha$, for $Q=3$ and $b=1$. In 
the inset are shown the optimal
overlap (solid lines) and activity (dashed lines) for $T=0$, $T=0.5$
and $T=1$; the unstable solutions for $m$ and $a$ are in light lines. In 
$\rm R_{1(2)}$ the strong retriever is a narrow (wide) retriever. NR is the
non--retriever phase. The dash--dotted line is the de~Almeida--Thouless line.
\\
\noindent
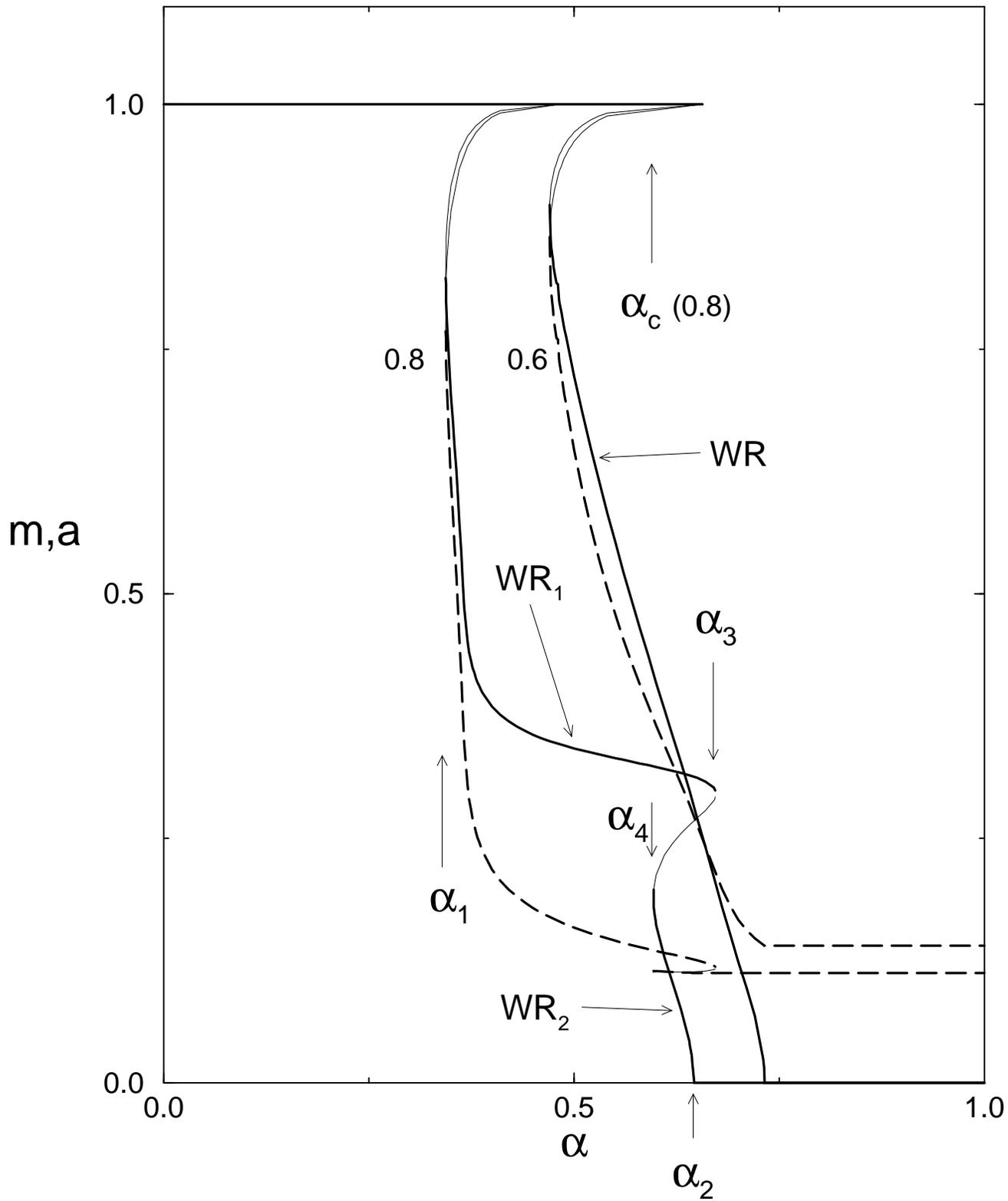
{\bf Figure 4:} Optimal overlap (solid lines) and activity (dashed
lines) for
$Q=4$, at $T=0$, for $b=0.6$ and $b=0.8$. The various $\alpha$
indicate the loads for which the optimal solutions appear or disappear,
for
$b=0.8$, and WR, ${\rm WR}_1$ and ${\rm WR}_2$ are weak retrievers.
\\
\noindent
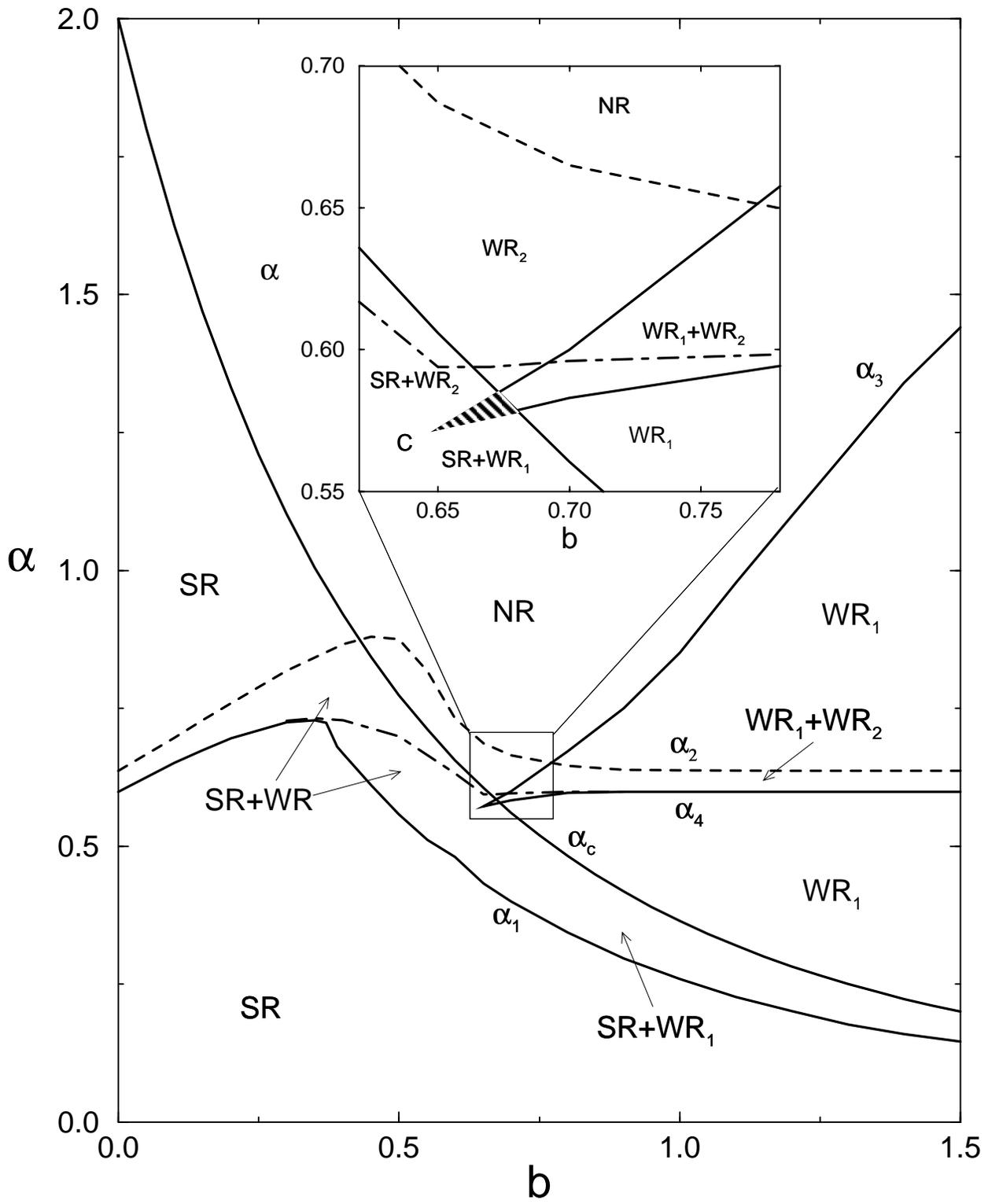
{\bf Figure 5:} Phase diagram for $\alpha$ as a function of $b$, for
$Q=4$ at $T=0$, described in the text. The amplified central part is shown 
separately. The retrievers and the nature (continuous or discontinuous) of 
the phase boundaries are as in previous figures.
The SR, ${\rm WR}_1$ and ${\rm WR}_2$ coexist in the shaded area of the
inset. The de~Almeida--Thouless line is the dash--dotted line.
\\
%

\end{document}